\documentclass[a4paper,11pt,article,oneside]{memoir}
\usepackage{ecis2019}

\usepackage{graphicx} 
\usepackage{csquotes}
\usepackage{enumitem}
\usepackage[center]{caption}
\usepackage[section]{placeins}

\maintitle{EEG-Based Detection of Braking Intention During Simulated Driving}

\authors{
Xinbin Liang, National University of Defense Technology, Changsha, China, lxb2203@163.com

Yang Yu, National University of Defense Technology, Changsha, China, yuyangnudt@hotmail.com

Yadong Liu, National University of Defense Technology, Changsha, China, liuyadong1977@163.com

Kaixuan Liu, National University of Defense Technology, Changsha, China, 2314825269@qq.com

Yaru Liu, National University of Defense Technology, Changsha, China, lyrnudt@163.com

Zongtan Zhou, National University of Defense Technology, Changsha, China, narcz@163.com
}

\shortauthors{Xinbin Liang, et al.} 

\usepackage{amsmath}

\begin{document}
\vspace{-1.5cm}
\begin{center}
\end{center}
\vspace{1cm}
\begin{abstract}\noindent
Accurately detecting and identifying drivers' braking intention is the basis of man-machine driving. In this paper, we proposed an electroencephalographic (EEG)-based braking intention measurement strategy. We used the Car Learning to Act (Carla) platform to build the simulated driving environment. 11 subjects participated in our study, and each subject drove a simulated vehicle to complete emergency braking and normal braking tasks. We compared the EEG topographic maps in different braking situations and used three different classifiers to predict the subjects' braking intention through EEG signals. The experimental results showed that the average response time of subjects in emergency braking was 762 ms; emergency braking and no braking can be well distinguished, while normal braking and no braking were not easy to be classified; for the two different types of braking, emergency braking and normal braking had obvious differences in EEG topographic maps, and the classification results also showed that the two were highly distinguishable. This study provides a user-centered driver-assistance system and a good framework to combine with advanced shared control algorithms, which has the potential to be applied to achieve a more friendly interaction between the driver and vehicle in real driving environment.
\end{abstract}


\chapter{Introduction}

Recently, the application of EEG in the field of driving safety has attracted great attention. These studies can be categorized as drivers’ fatigue, distraction and intention detection. Drivers’ intention studies were mostly about braking intention, and the EEG-based research received the most extensive attention  (\cite{1,2,3,4,5,6,9,10,11}). Haufe et al used event-related potentials (ERP) to predict the upcoming emergency braking situations (\cite{2}). The experiment was carried out on a driving simulator. Based on EEG signals, the occurrence of emergency braking can be predicted 130ms before the driver’s real braking action. Subsequently, Haufe et al carried out the above experiment in real-world driving, and the experimental results resembled those during simulated driving (\cite{3}). The experimental result verified the feasibility of predicting emergency braking intention through EEG signals. Kim et al investigated neural correlates of braking under various braking situations in simulated driving environment, and results showed that the braking intention could be detected by EEG in different scenarios (\cite{5}). Teng et al proposed a new method to detect the intention of emergency braking by using EEG signals (\cite{6}). Their proposed model combined regularized linear discriminant analysis (RLDA) with spatial-frequency features. 12 subjects participated in the experiment, and results showed the effectiveness of the proposed method. Hernandez et al investigated the feasibility of using driver’s EEG signals to recognize the intention of emergency braking when a driver experienced cognitive states such as workload, fatigue, and stress. The average recognition accuracy of emergency braking intention was more than 70\% in three different cognitive states (\cite{4}). Bi et al improved the accuracy of detecting the driver's emergency braking intention by combining EEG signals and the external environment of the vehicle, which provided a new idea for man-machine driving (\cite{11}). The driver's braking includes not only emergency braking (also known as sharp braking), but also normal braking (also known as soft braking), which is very common during driving. However, the studies on EEG-based detection of braking intention mainly focused on the field of emergency braking, little attention was paid to the detection of the driver's normal braking intention. If different types of braking can be identified before the real braking action, the vehicle will know the driver's specific braking intention, so as to take some measures in advance to achieve a more natural and effective human vehicle interaction.

In this paper, we mainly study whether the driver's two different types of braking intention (emergency braking and normal braking) can be recognized in advance by EEG, and whether the two can be effectively distinguished. For the sake of safety, we use the simulated driving platform to conduct the experiment. Relevant research showed that in terms of emergency braking intention detection, the simulated driving environment had similar results to the real driving environment (\cite{3}). The driving platform simulation adopted Carla, which is an open urban driving simulator (\citet{7}). In the experiment, subjects drove a simulated vehicle to complete a series of emergency braking and normal braking tasks during driving. In emergency braking situations, the subjects needed to make emergency braking immediately according to the external clues. And in normal braking situations, subjects spontaneously completed the braking action without external clues. The EEG signals of the subjects were recorded synchronously. Then we used the classification algorithms to identify the subject’s braking intention and judged the feasibility of the system.

The rest of this paper was organized as follows. The Methods and Materials section described subjects, experimental setup, data collection, classification algorithm, and performance metrics. The Results section showed the experimental results. In the end, we discussed and concluded this study, and described the possible application scenarios of our work.

\chapter{MATERIALS AND METHODS}

\section{Subjects}
A total of 11 subjects (aged 22-36 years old, with an average age of 25.73 years old, 9 males and 2 females) participated in the experiment. All subjects were recruited from school volunteers, obtained driving licenses and had more than 2 years of driving experience. Each subject was right-handed and had normal or corrected-to-normal vision. All subjects reported no history of mental diseases or other neurological diseases. Before the experiment, we explained the purpose and the procedure of the experiment to each subject, and all subjects participating in our study  wrote informed consent according to the Declaration of Helsinki. Each subject had enough sleep (>=8 hours) before the experiment, and did not take any drugs within three days before the experiment. During the experiment, each subject could end the experimental task at any time without any punishment. If the subject successfully completed the experiment, he/she would be paid 400RMB.

\section{Experimental Setup}
\begin{figure}[t]
\begin{center}
  \includegraphics[width=\textwidth]{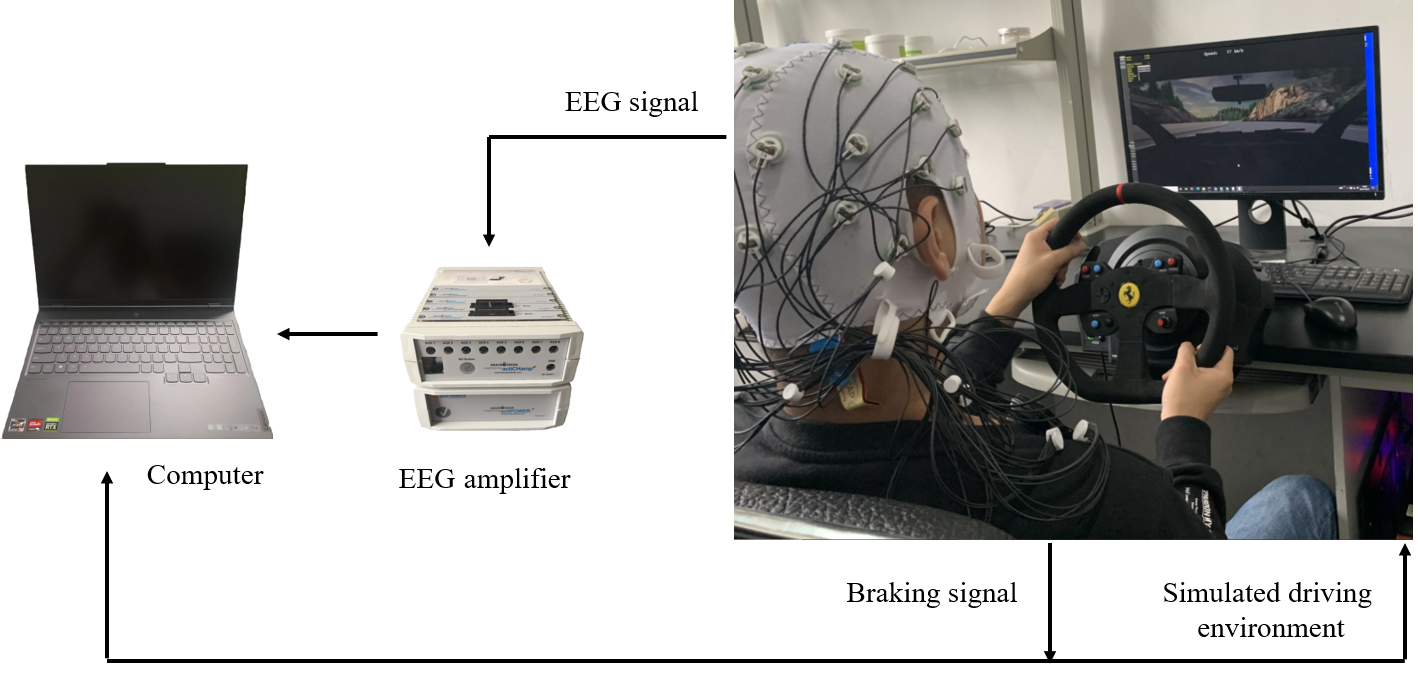}
  \caption{An illustration of the experimental platform.}
  \label{fig:Fig_1}
\end{center}
\end{figure}

As shown in Figure \ref{fig:Fig_1}, our experimental platform consisted of a driving simulator (composed of a driver’s seat, a steering wheel, a gas pedal and a brake pedal), an EEG acquisition equipment (ActiCHamp, Brain products, Germany) and two computers. The EEG acquisition equipment would record the subject’s scalp EEG signals during driving. By using the application program interface (API) of Carla driving simulator, we realized the automatic labeling of the EEG signals. The computers had two functions: (1) as the user interface, presented the driving simulation environment; (2) as the recording device, recorded the EEG signals with labels in real time.

\begin{figure}[t]
\begin{center}
  \includegraphics[width=\textwidth]{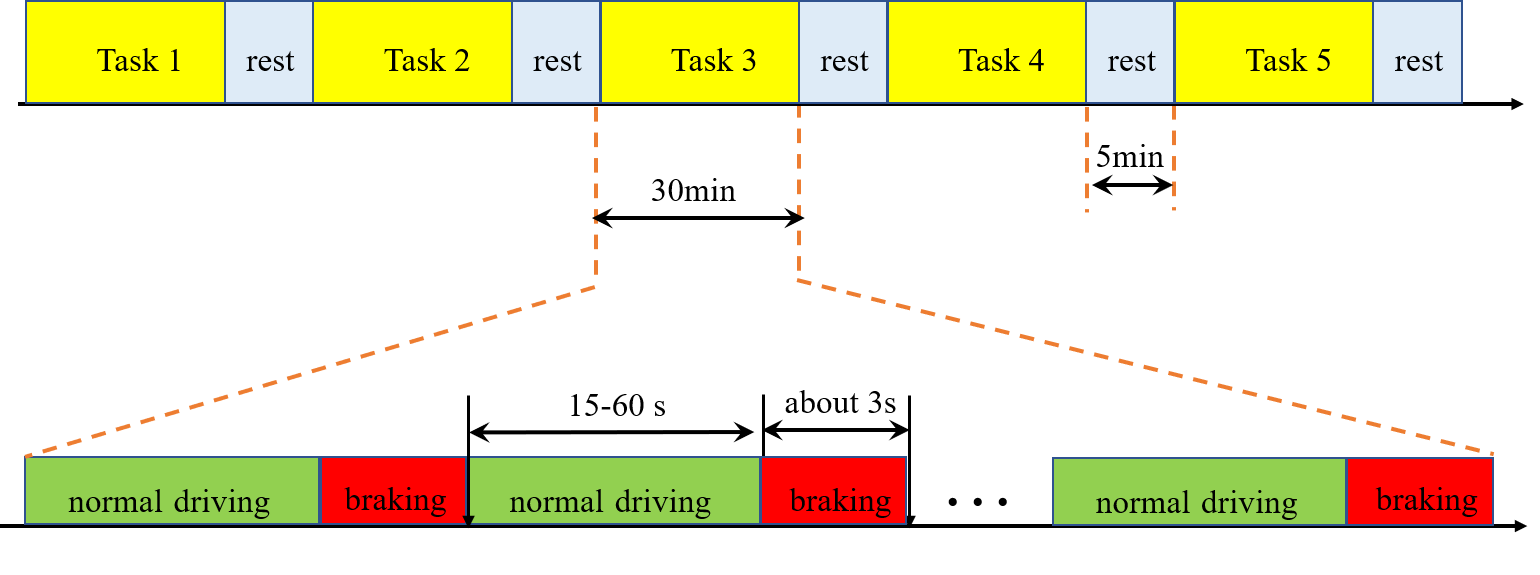}
  \caption{Timing scheme of the experimental paradigm.}
  \label{fig:Fig_2}
\end{center}
\end{figure}

The experiment included two different types of braking situations, one was emergency braking and the other was normal braking. In emergency braking situations, the setup consisted of two virtual vehicles, the vehicle in front (also known as lead vehicle) and the following vehicle. The lead vehicle was controlled by the instructor and the speed was kept at 90 km/h. The following vehicle was controlled by the subject, driving in the same lane, and keeping a distance of 6-12m from the lead vehicle. The lead vehicle in front would decelerate rapidly at random (the time interval varied from 15s to 60s),  and reminded the following vehicle through the brake lights. In order to avoid collision, the subject needed to make emergency braking immediately when he/she saw the front vehicle’s brake lights started flashing. In this process, two moments were recorded through the API of Carla platform, one was the time when the brake lights of the lead vehicle were on, and the other was the time when the subject of the following vehicle stepped on the brake pedal. 3 seconds after the deceleration, the lead vehicle accelerated again to 90 km/h. The subject continued to drive the vehicle to follow the lead vehicle, and kept the distance between 6m and 12m again. The distance between the two vehicles was displayed on the window of the simulation platform, and the subject could see it in real time. In normal braking situations, the experimental setup was similar to that of the emergency braking, except that the lead vehicle always maintained a speed of 90 km/h and did not decelerate randomly. The subject drove the same vehicle on the same road as in the emergency braking situations, and broke the every 15-60 s spontaneously and randomly. After depressing the brake pedal for about 3s, the subject accelerated the vehicle to follow the lead vehicle and kept the distance at 6-12m. Each subject needed to complete 5 driving tasks, and each lasting for 30 minutes. After each driving task, the subject rested for 5 minutes, as shown in Figure \ref{fig:Fig_2}.

\section{Data Acquisition}
In this study, we used the ActiCHamp amplifier and its electrodes to record the subject’s EEG signals. The electrode arrangement adopted the internationally accepted 10-20 system, in which 28 electrodes (F5 F3 Fz F4 F6 FT7 FC5 FC1 FC2 FC6 FT8 T7 C3 Cz C4 T8 CP5 CP1 CP2 CP6 P5 P3 Pz P4 P6 O1 Oz and O2) were used to record data, 2 electrodes (TP9 and TP10) were used as reference electrodes, and 1 electrode (FPz) was the grounding electrode. The EEG data acquisition sampling rate was set to 200Hz. We only carried out a simple preprocessing of the EEG data in this paper. First, a band-pass filter of FIR with a low-frequency of 1Hz and a high-frequency of 45Hz was used to filter the original signals. Then, the processed signals with an amplitude greater than 300µV were eliminated. The EEG signals corresponded to three kinds of situations: emergency braking, normal braking and no braking. In emergency braking and normal braking situations, we selected the target epochs ranged from -3000ms to 1000ms relative to the moment when the subject started pressing the brake pedal. The no braking epochs were extracted by a sliding window (4000ms length) on the EEG signals that were at least 3000ms away from any subjects’ braking behavior. Baseline correction was performed epoch-wise by subtracting the average EEG signals in the first 500ms of the epoch. In total, we got 189±54 epochs in emergency braking and 114±27 epochs in normal braking per subject, respectively. In addition, we randomly selected 200 epochs of no braking data from each subject through the sliding window.

\section{Classification Algorithms}
We selected three commonly used representative algorithms in the field of brain computer interface (BCI) for EEG classification, which are Common Spatial Pattern filter combined with Linear Discriminant Analysis (CSP+LDA) (\cite{12}), Riemannian Minimum Distance to Mean (RMDM) (\cite{14}), and the EEGNet algorithm based on deep learning (\cite{8}). As a traditional EEG classification algorithm, CSP+LDA algorithm has very good performance and is widely used in different types of BCI (\cite{13}). Algorithms based on Riemannian Geometry have changed some of the conventions adopted in the traditional approaches; instead of estimating the optimal filters, Riemannian Geometry classifiers (RGCs) map the EEG data directly onto a geometrical space, which has a suitable metric for classification. Despite their short time in EEG decoding, RGCs have received extensive attention in BCI, including the winning score obtained in five recent international BCI competitions (\cite{13}). Using the training data, RMDM classifier computes a geometric mean for each class, and then assigns test data to the class corresponding to the closest mean. RMDM is a simple yet efficient classifier,  and has been reported to result in a high classification accuracy (\cite{15}). EEGNet is a compact convolutional neural network specially designed for EEG classification. It has the advantages of simple structure, short training time and high classification accuracy. More importantly, EEGNet is an end-to-end classification method, which can be conveniently used for online classification (\cite{8,16}).

There were three categories: emergency braking, normal braking and no braking. We conducted comparative classification studies in three different combinations, namely, emergency braking vs. no braking, normal braking vs. no braking, emergency braking vs. normal braking. For each subject, we trained the classifier respectively. The number of epochs  in the three categories was not equal. We chose epochs of the other two categories based on the category with the least epochs, so that each subject had the same number of epochs in the three categories. The length of the EEG data input to the classifier was 1000ms. We used the EEG data of 1000ms before braking to train the classifier, and adopted a sliding window with a length of 1000ms to get the data for testing. For each subject, a total of 10 times of training and testing were carried out. Each time, we randomly selected 50\% of the extracted epochs as the training set and the rest as the test set. The classifier was trained on the training set and its performance was evaluated on the test set. We calculated the average of 10 times as the final result. 

\section{Performance Metrics}
We used the prediction time and its corresponding classification accuracy to measure the performance of the system. The classification accuracy rate p was defined as the ratio of correct classification number Nc to the total amount of data N, as shown in \eqref{eq}. The prediction time was defined as how long the classifier can recognize the subject’s braking intention at a certain accuracy before the subject stepped on the brake pedal.

\begin{equation}
p=Nc/N\label{eq}
\end{equation}

\chapter{RESULTS}

\section{Emergency braking response time}
Emergency braking response time (EBRT) was the duration from the moment when the lead vehicle’s brake lights started flashing to the moment when the subject started to step on the brake pedal under an emergency braking situation. In our experiment, the average EBRT was 762±156ms, the maximum was 1490ms, and the minimum was 300ms. The distribution of the EBRT was skewed with percentiles P5=520ms, P25=660ms, P50=750ms, P75=850ms, and P95=1020ms.

\section{The EEG Topographic Map}
\begin{figure}[t]
\begin{center}
  \includegraphics[width=\textwidth]{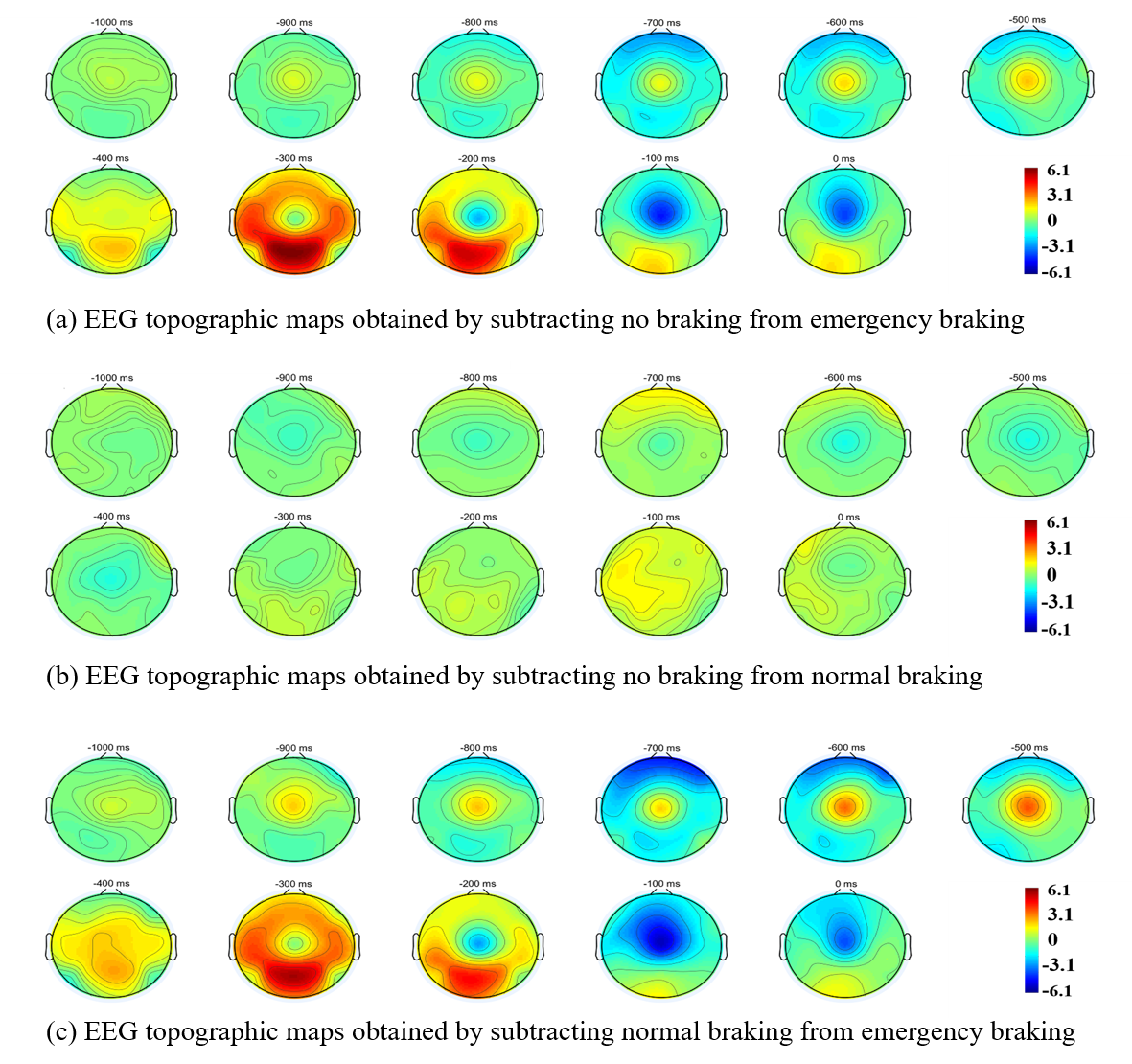}
  \caption{EEG topographic maps. -1000ms represents 1000 milliseconds before the subject started braking, and 0ms is the onset of braking. The warmer color indicates the higher value of the EEG potential.}
  \label{fig:Fig_3}
\end{center}
\end{figure}

Figure \ref{fig:Fig_3} showed the EEG topographic maps with an interval of 100ms from 1000ms before braking to the onset of braking. Figure \ref{fig:Fig_3} (a)  and Figure \ref{fig:Fig_3} (b) were obtained by subtracting the no braking epochs from emergency braking and normal braking respectively. Figure \ref{fig:Fig_3} (c) was obtained by subtracting normal braking epochs from emergency braking. 0ms represented the onset of braking. The warmer the color, the higher the potential of EEG signals. From 400ms before emergency braking, the EEG potential started to rise significantly in the occipital area, and the maximum potential was approximately 6µV reached at about 300ms before subjects’ braking action. We believe that this potential is similar to P300, which was also induced by the random flashing stimulation similar to the oddball paradigm. From 200 milliseconds before braking, the central region of the brain (mainly responsible for movement) began to have a negative potential offset, which showing the subject's motor preparation potential for emergency braking. Compared with emergency braking, the EEG changes of normal braking were relatively flat. From 900ms to 400ms before normal braking, there was a small negative potential offset in the central region of the brain. No significant potential changes were found in the occipital region, which is mainly responsible for the processing of visual information. A comparison of the EEG topographic maps under emergency braking and normal braking was also conducted, as shown in Figure \ref{fig:Fig_3} (c) . We can see that obvious differences in EEG potential between the two braking modes, indicating that emergency braking and normal braking are  significantly different in cognitive processing. This also provides some physiological basis for classifying emergency braking from normal braking.

\section{Classification Performance}
Figure \ref{fig:Fig_4} showed three sets of classification accuracy curves for detecting two different types of braking intention. Figure \ref{fig:Fig_4} (a) showed the average classification accuracy curves of emergency braking vs. no braking, Figure \ref{fig:Fig_4} (b) showed the average classification accuracy curves of normal braking vs. no braking, and Figure \ref{fig:Fig_4} (c) showed the average classification accuracy curves of emergency braking vs. normal braking. Three different classification algorithms, CSP+LDA, RMDM and EEGNet, were used respectively. The EEG data of subjects 1000ms before braking were used to train classifiers. The test data was obtained through a sliding window with a length of 1000ms, and the accuracy of classification corresponded to the end point of the window. The data of each subject were trained and tested respectively, and the average values of all subjects were taken to generate the accuracy curves as shown in Figure \ref{fig:Fig_4}. We got the following results:

\begin{figure}
\begin{center}
  \includegraphics[width=\textwidth]{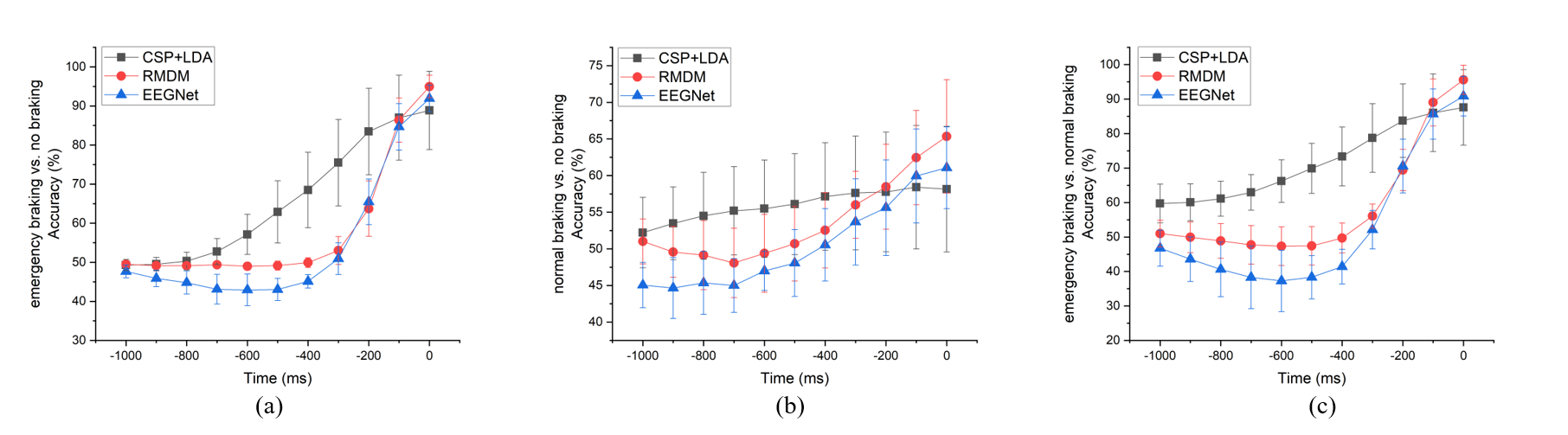}
  \caption{Classification accuracy curves for detecting braking intention. 0 on the Time axis represents the onset of braking. The results were obtained through a sliding window with a length of 1000ms, and the accuracy of classification corresponds to the end point of the window.}
  \label{fig:Fig_4}
\end{center}
\end{figure}

\begin{itemize}
    \item Emergency braking and no braking can be well distinguished. At the braking moment, the average accuracy of CSP+LDA, RMDM and EEGNet algorithms were 88.85±10.02\%, 94.97±2.85\% and 91.91±3.39\% respectively; 100 milliseconds before braking action, the average accuracy were 86.99±10.89\%, 86.39±5.64\% and 84.66±5.96\% respectively.
    \item The classification results of normal braking vs. no braking were poor. At the moment of braking, the average classification accuracy of CSP+LDA, RMDM and EEGNet algorithms were 58.14±8.57\%, 65.35±7.74\% and 61.06±5.58\% respectively; 100 milliseconds before braking action, the classification accuracy were 58.42±8.43\%, 62.45±6.43\% and 59.94±6.40\% respectively.
    \item Two different braking modes (emergency braking and normal braking) can be effectively distinguished. At the braking moment, the average classification accuracy of CSP+LDA, RMDM and EEGNet algorithms were 87.57±10.96\%, 95.59±4.25\% and 90.84±5.74\% respectively; 100 milliseconds before braking, the accuracy were 86.01±11.26\%, 89.03±6.81\% and 85.64±7.29\% respectively. 
    \item When the test data selected by the sliding window were consistent with the training data  in time, the RMDM based on Riemannian Geometry achieved the higher average  classification accuracy than other algorithms in all three classification cases. When the test data selected by the sliding window were not inconsistent with the training data in time, such as when the test data was 200ms ahead of the training data, CSP+LDA algorithm had a comparative advantage. EEGNet is a classifier based on deep learning, which requires a large number of samples for training. In this paper, we trained the EEG data of each subject separately, and the classification results of EEGNet was not as good as RMDM.
\end{itemize}

\chapter{DISCUSSION AND CONCLUSION}

In this paper, we mainly investigated whether the braking intention can be detected by the driver’s EEG signals while driving. We conducted experiments on a simulated driving platform and collected EEG data of 11 subjects. Two different types of braking intention were analyzed. We carried out EEG topographic maps for comparison, and used three classifiers to detect the subject’s braking intention. Through the analysis of EEG topographic maps, we found that compared with normal braking, the EEG changed significantly before emergency braking. The reason may be that under normal braking, the subjects took a spontaneous and relatively gentle action, while under emergency braking, the subjects had a sense of tension after the stimulation and needed to complete the braking behavior as soon as possible. The classification results also proved our conjecture that it was easier to detect the intention of emergency braking. At the moment of emergency braking, RMDM algorithm achieved the highest average classification accuracy of nearly 95\%. Even 200ms before emergency braking, CSP+LDA still achieved an average classification accuracy of about 83\%. For the detection of normal braking intention, the average classification accuracy of about 65\% can be achieved by RMDM at the moment of braking, but only about 60\% 100ms before braking. For the classification algorithm, the RMDM algorithm based on Riemannian Geometry had the best classification results, but CSP+LDA algorithm showed advantages in early braking intention detection. Due to the lack of a large number of training samples, the EEGNet algorithm based on deep learning did not achieved good classification results. The challenge of insufficient samples needs to be solved when using deep learning methods in the field of EEG classification (\cite{8,16}).

Our work mainly focuses on the recognition of emergency braking and normal braking, and provides a good shared control framework for users. When a user is faced with a sudden driving situation, he/she may make mistakes due to nervousness. Theoretically, our system can assist the user at this time to provide more guarantee for safe driving. In the future, we intend to complete the above experiments in real vehicles and real driving scenes, and focus on improving the reliability of the system.

\newpage

\printbibliography
\end{document}